\documentclass[12pt]{article}

\usepackage{graphicx}
\usepackage{amsmath,amssymb}
\usepackage{bm}
\usepackage{braket}
\usepackage{hyperref}
\usepackage{placeins}
\usepackage{adjustbox}

\title{Birth and Death of Entanglement in\\
Hamiltonian-Driven Quantum Games\\ under Decoherence}

\author{
Teena Thomas$^{1}$, S. Balakrishnan$^{1}$\\[4pt]
$^{1}$Department of Physics, School of Advanced Sciences,\\
Vellore Institute of Technology, Vellore 632014, Tamil Nadu, India\\
\texttt{physicsbalki@gmail.com}
}

\date{}

\begin{document}

\maketitle

\begin{abstract}
Quantum game theory investigates the influence of quantum resources on strategic decision-making. In this work, two-player quantum games based on the Transverse Field Ising Model (TFIM) are investigated under amplitude-damping decoherence. The TFIM Hamiltonian naturally produces a family of entangling gates, enabling a physically motivated implementation of quantum games. The effects of noise on Nash equilibria, players' payoffs, concurrence and coherence of the quantum states for different initial states and strategy pairs are analyzed. The results show that decoherence progressively suppresses quantum strategic advantages, with maximum damping driving all outcomes to identical classical payoffs. The concurrence and coherence analysis of the states generated in the quantum game reveal initial state and strategy dependent quantum correlation dynamics, including entanglement sudden birth and death.
\end{abstract}

\section{\label{sec:level1}Introduction}
Game theory provides a mathematical framework for analyzing strategic decision making among rational players and the payoff of each player depends on the strategies chosen by all players~\cite{nash1950equilibrium,rubinstein2007theory}. Among the various game-theoretic models, the Prisoner's Dilemma has become one of the most extensively studied frameworks as it captures the conflict between individual rationality and collective benefit. The Nash equilibrium predicts mutual defection as the rational outcome, although mutual cooperation yields a higher collective payoff. This dilemma has motivated extensive research into mechanisms capable of promoting cooperative behavior~\cite{benjamin2020emergence}.\\
An important development in this direction is the correspondence between classical game theory and the Ising model~\cite{galam2010ising}. By mapping the payoff matrix onto Ising interaction parameters, strategic behavior can be analyzed using methods from statistical physics, particularly in the thermodynamic limit where the collective behavior of a large population emerges naturally~\cite{sarkar2019quantum,adami2018thermodynamics}. This connection has established spin models as powerful tools for investigating cooperative phenomena and evolutionary games.\\
Extending this correspondence to the quantum domain allows players to exploit quantum mechanical resources such as superposition and entanglement, leading to strategic outcomes that are inaccessible in classical games~\cite{piotrowski2003invitation}. Consequently, quantum games have become an important platform for understanding the interplay between quantum mechanics and decision theory~\cite{meyer1999quantum,flitney2002introduction}.\\
Although most quantum game formulations employ abstract entangling operators~\cite{eisert1999quantum,marinatto2000quantum}, practical implementation requires a physical system capable of naturally generating quantum correlations. Quantum spin systems are particularly attractive in this regard as their Hamiltonians directly govern the evolution of interacting qubits~\cite{fan2005optimal}. Thus, the spin dynamics itself can serve as a mechanism for quantum game realization.\\
Among various spin Hamiltonians, the Transverse Field Ising Model (TFIM) provides one of the simplest, yet fundamentally important models~\cite{sachdev1999quantum,huang2016quantum}. It incorporates both nearest-neighbour spin interaction $J$ and an external transverse magnetic field. The time evolution $(t)$ generated by the TFIM Hamiltonian naturally produces a family of quantum gates whose entangling capability can be controlled through the evolution parameter which is an interaction-time value $(Jt)$. This makes TFIM an attractive physical platform for implementing quantum games.\\
However, real quantum systems are inevitably affected by environmental interactions. Decoherence degrades quantum coherence and entanglement, both of which are responsible for the quantum advantage in strategic games~\cite{flitney2005quantum}. Therefore, any physically realizable quantum game must be analyzed in the presence of noise to determine whether the strategic advantages in the ideal case remain robust. In this work, amplitude damping noise, which models irreversible energy dissipation from a quantum system to its environment  is studied. Thus, TFIM-driven quantum games under amplitude damping decoherence is analyzed. To be precise, the influence of noise on players' payoffs and Nash equilibria  for different initial states are examined. To understand how the observed strategic behavior correlates with the degradation of entanglement, the concurrence and coherence of the decohered state are analyzed further. This quantified analysis provides insight into the robustness of Hamiltonian-driven quantum games under realistic noisy conditions.\\
As the amplitude damping strength increases, entanglement may vanish completely beyond a critical noise level and may reappear within a limited evolution parameter regime. In this work, the parameter $(Jt)$ is closely examined to investigate the emergence and disappearance of entanglement under amplitude damping. Sudden birth~\cite{lopez2008sudden,xu2009sudden} of entanglement is observed when concurrence changes from zero to a nonzero value as $Jt$ varies, whereas entanglement sudden death~\cite{yu2005evolution,yu2006sudden,yu2009sudden,yang2013relation} is realized when concurrence subsequently vanishes over a finite $Jt$ interval. Investigating these noise-induced transitions in different initial states are essential for understanding the robustness of Hamiltonian-generated entanglement in open quantum games.\\
The paper is structured as follows. Section \ref{sec:level2} introduces the transverse-field Ising model and the corresponding interaction dynamics. Section \ref{sec:level3} presents the quantum game framework under amplitude damping, including the game formulation, the amplitude-damping channel, and the calculation of the payoffs for the noiseless, intermediate-noise, and maximum-noise cases. Section \ref{sec:level4} investigates the entanglement properties of the resulting quantum states. The concurrence measure is introduced, followed by an analysis of the entanglement dynamics for symmetric and asymmetric initial states, including the occurrence of entanglement sudden birth and death. The behavior of quantum coherence is also examined. Finally, Section \ref{sec:level5} discusses the main results and their implications.

\section{\label{sec:level2}Transverse Field Ising Model}

In this work, Transverse Field Ising Model (TFIM) is considered as the physical system responsible for generating the entangling operation used in the quantum game. Owing to its simple spin-spin interaction and controllable transverse field, the TFIM provides a natural Hamiltonian-based realization of two-qubit quantum dynamics.\\
The Hamiltonain of the TFIM is~\cite{sachdev1999quantum},
\begin{equation}
    H=-J\sigma_z^{(1)}\sigma_z^{(2)}-h(\sigma_x^{(1)}+\sigma_x^{(2)})
    \label{eq:TFIM}
\end{equation}
where $J$ is the interaction strength which tries to align the spins in the $z$ axis or the longitudinal direction, $h$ is the transverse magnetic field which tries to align the spins along the $x$ axis or the transverse direction and $\sigma_z$ and $\sigma_x$ are the pauli $z$ and pauli $x$ matrices respectively. Here, the superscripts $(1)$ and $(2)$ denote the spin on which the corresponding Pauli operator acts.\\
The unitary evolution operator generated by the TFIM Hamiltonian is obtained by diagonalizing the Hamiltonian and evaluating the exponential time-evolution operator. Thus,
\begin{equation}
    U=e^{-iHt}
\end{equation}
where $t$ is the evolution time.\\
A detailed derivation of the unitary operator in terms of the Hamiltonian parameters is presented in Ref.~\cite{Thomas2026StrategicEquilibria}. Expressing this unitary in the Cartan decomposition and identifying its nonlocal coordinates gives
    \begin{equation}
        (c_1,c_2,c_3)=(Jt,Jt,0)
        \label{eq:c1}
    \end{equation}
These coordinates place the TFIM evolution along the $OA_2$ edge of the Weyl chamber, corresponding to the $iSWAP$ family of nonlocal gates~\cite{makhlin2002nonlocal,rezakhani2004characterization,zhang2003geometric}. Consequently, the evolution generates a continuous family of two-qubit entangling operations as $Jt$ is varied. Thus, the entangling capability of the quantum game is determined by the evolution parameter $Jt$, which serves as the effective control parameter throughout this work. At $Jt=0$, the Cartan coordinates $(0,0,0)$ correspond to the local operation. At $Jt=\pi/4$, the coordinates $(\pi/4,\pi/4,0)$ correspond to the $\sqrt{iSWAP}$ gate, while $Jt=\pi/2$ corresponding to $(\pi/2,\pi/2,0)$, gives the $iSWAP$ gate.\\
This physically generated entangling operation is then employed as the entangling operator in the two-player quantum Prisoner's Dilemma considered in this work. Instead of introducing an arbitrary entangling gate, the game is constructed using the two-qubit evolution naturally generated by the TFIM. The resulting game is subsequently analyzed under amplitude-damping decoherence, with particular emphasis on its payoff structure and entanglement dynamics.
\FloatBarrier

\section{\label{sec:level3}Quantum Game under Amplitude Damping}

\subsection{Quantum Game Framework}

\begin{figure}[!h]
\includegraphics[width=\columnwidth]{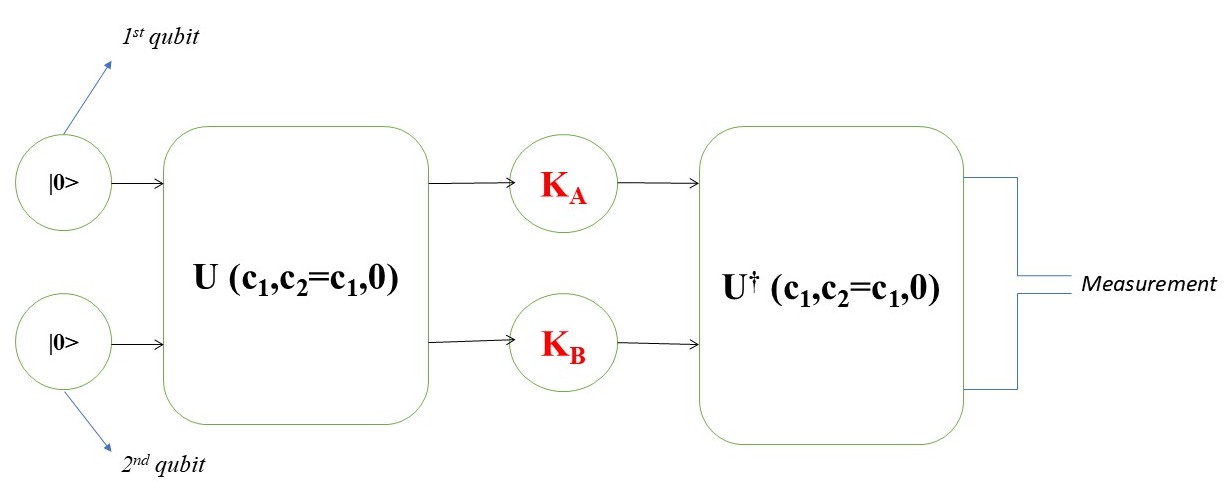}
    \caption{Schematic representation of the two-qubit quantum game protocol.}
    \label{fig:mewl}
\end{figure}

The unitary evolution generated by the TFIM Hamiltonian is incorporated into the modified Eisert–Wilkens–Lewenstein (EWL) framework \cite{vijayakrishnan2019role} to realize a Hamiltonian-driven quantum Prisoner's Dilemma. The game begins with an initial two-qubit state as shown in FIG.~\ref{fig:mewl}.
\\The TFIM evolution operator $U$ acts as the entangling operation. $Player_1$ and $Player_2$ subsequently apply their classical strategies $K_A$ and $K_B$, chosen from the set ({$I$,$\sigma_x$}). Finally, the disentangling operator $U^\dagger$ is applied before measurement, and the resulting density matrix is used to evaluate the players' payoffs.\\
In the present work, each player is allowed to choose one of the two classical strategies: the Identity operator $I$, representing cooperation ($C$) and the Pauli-$x$ operator $\sigma_x$ representing defection ($D$) in the Prisoner's Dilemma game setting.
\FloatBarrier

\subsection{Amplitude Damping}

Since realistic quantum systems inevitably interact with environment, the quantum game is analyzed in the presence of amplitude damping, one of the commonly considered noise models. This noisy channel models energy dissipation from the qubits to the environment through spontaneous decay from the excited state to the ground state, leading to the degradation of quantum coherence and entanglement.\\
The single qubit Kraus operators for amplitude damping are~\cite{nielsen2000quantum},
\begin{equation}
    M_0 =
\begin{pmatrix}
1 & 0 \\
0 & \sqrt{1-p}
\end{pmatrix},
\qquad
M_1 =
\begin{pmatrix}
0 & \sqrt{p} \\
0 & 0
\end{pmatrix}
\label{eq:amplitude_damping}
\end{equation}
where $p$ is the damping probability, with $p=0$ representing the noiseless case and $p=1$ corresponding to complete amplitude damping.
\begin{figure}
\includegraphics[width=\columnwidth]{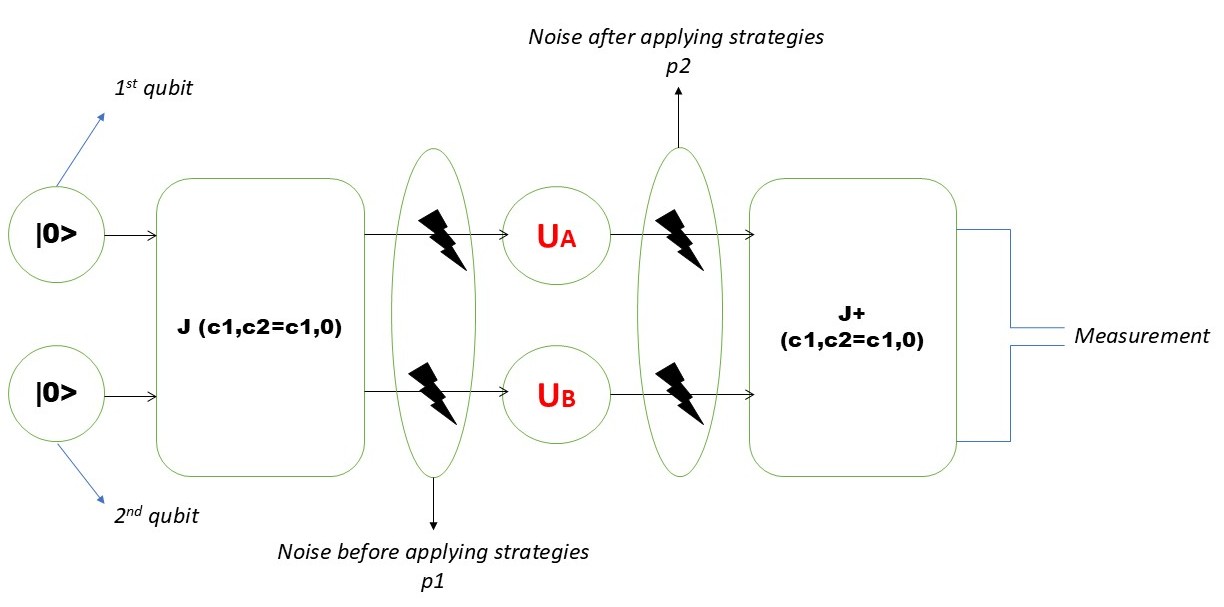}
    \caption{Schematic representation of the two-qubit quantum game protocol with amplitude-damping noise applied before and after the players' strategic operations.}
    \label{fig:figure1}
\end{figure}
Considering FIG.~\ref{fig:figure1}, the protocol begins with the initial two-qubit state $\ket{00}$, which is evolved under the Hamiltonian-generated entangling operator $U(c_1=c_2=Jt,c_3=0)$. Amplitude-damping noise with strength $p_1$ is then applied to the two qubits before the players implement their respective unitary strategies $K_A$ and $K_B$. Subsequently, a second amplitude-damping channel with strength $p_2$ is introduced after the application of the strategic operations. The resulting state is then subjected to the disentangling operation $U^\dagger(c_1=c_2=Jt,c_3=0)$, followed by measurement in the computational basis to obtain the outcome probabilities and corresponding payoffs. In this work, the noise before and after applying strategies are taken as same. Thus, $p_1=p_2=p$ throughout the analysis for the sake of mathematical simplicity.
\FloatBarrier

\subsection{Payoff Calculation}
Following the density matrix formalism presented in Ref.~\cite{chen2003quantum}, the expected payoff for the quantum Prisoner's Dilemma is calculated as,
\begin{equation}
\langle \$ \rangle
=
3P_{CC}+P_{DD}+5P_{DC}+0P_{CD}
\label{eq:payoff}
\end{equation}
where the probabilities are obtained from the final density matrix $\rho_f$ as,
\begin{equation}
P_{ij}
=
\operatorname{Tr}
\left(
\rho_f\ket{i}\bra{j}
\right)
\label{eq:probability}
\end{equation}
Here, $P_{CC}$, $P_{CD}$, $P_{DC}$, and $P_{DD}$ denote the probabilities corresponding to the strategy pairs $(C,C)$, $(C,D)$, $(D,C)$, and $(D,D)$, respectively.

\subsubsection{Noiseless case $(p=0):$}

The payoff matrices for the different initial states and values of the
evolution parameter $Jt$ for the noiseless case are shown in TABLE~\ref{tab:payoff}.
\begin{table}
\caption{Payoff matrices for different initial states and values of the evolution parameter $Jt$ for $p=0$. Bold entries indicate Nash equilibrium outcomes. $A$ and $B$ denote Players $1$ and $2$ respectively.}
\centering
\begin{tabular}{@{}c|ccc|ccc|ccc}
\hline
\shortstack{Initial\\State}
&
\multicolumn{3}{c|}{\(Jt=0\)}
&
\multicolumn{3}{c|}{\(Jt=\pi/4\)}
&
\multicolumn{3}{c}{\(Jt=\pi/2\)}
\\
\hline
&
$A\backslash B$ & $I$ & $\sigma_x$
&
$A\backslash B$ & $I$ & $\sigma_x$
&
$A\backslash B$ & $I$ & $\sigma_x$
\\
%|00> 
$\ket{00}$
&
$I$ & $(3,3)$ & $(0,5)$
&
$I$ & {\textbf{(3,3)}} & $(2.5,2.5)$
&
$I$ & {\textbf{(3,3)}} & $(5,0)$
\\
&
$\sigma_x$ & $(5,0)$ & {\textbf{(1,1)}}
&
$\sigma_x$ & $(2.5,2.5)$ & $(1,1)$
&
$\sigma_x$ & $(0,5)$ & $(1,1)$
\\
\hline
%|11> 
$\ket{11}$
&
$A\backslash B$ & $I$ & $\sigma_x$
&
$A\backslash B$ & $I$ & $\sigma_x$
&
$A\backslash B$ & $I$ & $\sigma_x$
\\
&
$I$ & {\textbf{(1,1)}} & $(5,0)$
&
$I$ & $(1,1)$ & $(2.5,2.5)$
&
$I$ & $(1,1)$ & $(0,5)$
\\
&
$\sigma_x$ & $(0,5)$ & $(3,3)$
&
$\sigma_x$ & $(2.5,2.5)$ & {\textbf{(3,3)}}
&
$\sigma_x$ & $(5,0)$ & {\textbf{(3,3)}}
\\
\hline
%|01>
$\ket{01}$
&
$A\backslash B$ & $I$ & $\sigma_x$
&
$A\backslash B$ & $I$ & $\sigma_x$
&
$A\backslash B$ & $I$ & $\sigma_x$
\\
&
$I$ & $(0,5)$ & $(3,3)$
&
$I$ & $(0,5)$ & $(2,2)$
&
$I$ & $(0,5)$ & $(1,1)$
\\
&
$\sigma_x$ & {\textbf{(1,1)}} & $(5,0)$
&
$\sigma_x$ & {\textbf{(2,2)}} & $(5,0)$
&
$\sigma_x$ & {\textbf{(3,3)}} & $(5,0)$
\\
\hline
%|10> 
$\ket{10}$
&
$A\backslash B$ & $I$ & $\sigma_x$
&
$A\backslash B$ & $I$ & $\sigma_x$
&
$A\backslash B$ & $I$ & $\sigma_x$
\\
&
$I$ & $(5,0)$ & {\textbf{(1,1)}}
&
$I$ & $(5,0)$ & {\textbf{(2,2)}}
&
$I$ & $(5,0)$ & {\textbf{(3,3)}}
\\
&
$\sigma_x$ & $(3,3)$ & $(0,5)$
&
$\sigma_x$ & $(2,2)$ & $(0,5)$
&
$\sigma_x$ & $(1,1)$ & $(0,5)$
\\
\hline
\end{tabular}
\label{tab:payoff}
\end{table}
It presents the payoff matrices for the four computational-basis initial states at $Jt=0$, $Jt=\pi/4$, and $Jt=\pi/2$. The results reveal that the strategic outcomes depend on both the initial state and the evolution parameter $Jt$. Starting from $Jt=0$, the game reduces to the corresponding unentangled limit, while the Hamiltonian-driven evolution modifies the payoff structure as $Jt$ increases.  The comparison across the different initial states further demonstrates that the same Hamiltonian evolution can produce distinct strategic outcomes depending on the initial preparation of the two-qubit system. Therefore, the evolution parameter 
$Jt$ provides a direct control over the payoff structure and, consequently, the equilibrium behavior of the quantum game. The symmetrical initial states $\ket{00}$ and $\ket{11}$ give the maximum cooperative payoff $(3,3)$ as the Nash equilibrium at $Jt=\pi/4$ itself, whereas initial states $\ket{01}$ and $\ket{10}$ produce intermediate payoff value $(2,2)$ only. For the symmetrical initial states ($\ket{00}$ and $\ket{11}$), as $Jt$ evolves, Nash equilibrium strategy also evolves whereas for asymmetrical initial states ($\ket{01}$ and $\ket{10}$), Nash equilibrium strategy remains constant throughout the evolution.

\subsubsection{Intermediate noise case $(p=0.5):$}
For the intermediate noise case, the payoff matrices corresponding to the different initial states and the values of the evolution parameter $Jt$ are summarized in TABLE~\ref{tab:payoff2}.
\begin{table}
\caption{Payoff matrices for different initial states and values of the evolution parameter $Jt$ for $p=0.5$}
\scriptsize
\setlength{\tabcolsep}{2pt}
\begin{tabular}
{@{}c|ccc|ccc|ccc}
\hline
\shortstack{Initial\\State}
&
\multicolumn{3}{c|}{\(Jt=0\)}
&
\multicolumn{3}{c|}{\(Jt=\pi/4\)}
&
\multicolumn{3}{c}{\(Jt=\pi/2\)}
\\
\hline
&
$A\backslash B$ & $I$ & $\sigma_x$
&
$A\backslash B$ & $I$ & $\sigma_x$
&
$A\backslash B$ & $I$ & $\sigma_x$
\\
% |00>
$\ket{00}$
&
$I$ & $(3,3)$ & $(1.5,4)$
&
$I$ & {\textbf{(3,3)}} & $(2.75,2.75)$
&
$I$ & {\textbf{(3,3)}} & $(4,1.5)$
\\
&
$\sigma_x$ & $(4,1.5)$ & {\textbf{(2.25,2.25)}}
&
$\sigma_x$ & $(2.75,2.75)$ & $(2.25,2.25)$
&
$\sigma_x$ & $(1.5,4)$ & $(2.25,2.25)$
\\
\hline
% |11>
$\ket{11}$
&
$A\backslash B$ & $I$ & $\sigma_x$
&
$A\backslash B$ & $I$ & $\sigma_x$
&
$A\backslash B$ & $I$ & $\sigma_x$
\\
&
$I$ & {\textbf{(2.687,2.687)}}
& {\textbf{(2.687,2.687)}}
&
$I$ & {\textbf{(2.687,2.687)}}
& {\textbf{(2.687,2.687)}}
&
$I$ & {\textbf{(2.687,2.687)}}
& {\textbf{(2.687,2.687)}}
\\
&
$\sigma_x$ & {\textbf{(2.69,2.69)}}
& {\textbf{(2.69,2.69)}}
&
$\sigma_x$ & {\textbf{(2.687,2.687)}}
& {\textbf{(2.687,2.687)}}
&
$\sigma_x$ & {\textbf{(2.687,2.687)}}
& {\textbf{(2.687,2.687)}}
\\
\hline
% |01>
$\ket{01}$
&
$A\backslash B$ & $I$ & $\sigma_x$
&
$A\backslash B$ & $I$ & $\sigma_x$
&
$A\backslash B$ & $I$ & $\sigma_x$
\\
&
$I$ & $(2.25,3.5)$ & $(2.25,3.5)$
&
$I$ & $(2.25,3.5)$ & $(2.687,2.687)$
&
$I$ & {\textbf{(2.25,3.5)}} & $(3.125,1.875)$
\\
&
$\sigma_x$ & {\textbf{(3.125,1.875)}}
& {\textbf{(3.125,1.875)}}
&
$\sigma_x$ & {\textbf{(2.687,2.687)}}
& $(3.125,1.875)$
&
$\sigma_x$ & {\textbf{(2.25,3.5)}}
& $(3.125,1.875)$
\\
\hline
% |10>
$\ket{10}$
&
$A\backslash B$ & $I$ & $\sigma_x$
&
$A\backslash B$ & $I$ & $\sigma_x$
&
$A\backslash B$ & $I$ & $\sigma_x$
\\
&
$I$ & $(3.5,2.25)$ & {\textbf{(1.875,3.125)}}
&
$I$ & $(3.5,2.25)$ & {\textbf{(2.687,2.687)}}
&
$I$ & {\textbf{(3.5,2.25)}} & {\textbf{(3.5,2.25)}}
\\
&
$\sigma_x$ & $(3.5,2.25)$ & {\textbf{(1.875,3.125)}}
&
$\sigma_x$ & $(2.687,2.687)$ & $(1.875,3.125)$
&
$\sigma_x$ & $(1.875,3.125)$ & $(1.875,3.125)$
\\
\hline
\end{tabular}
\label{tab:payoff2}
\end{table}

Under intermediate amplitude damping, the dependence of the quantum Prisoner's Dilemma on the initial state remains evident despite the presence of decoherence. Although the noise reduces the quantum advantage, it does not eliminate the influence of the TFIM evolution or the distinct payoff characteristics of different initial states. The $\ket{00}$ state largely preserves its qualitative behavior observed in the noiseless case, whereas the $\ket{11}$  state exhibits identical payoffs for all strategy combinations and TFIM evolution parameters, indicating that intermediate noise suppresses its strategic sensitivity. In contrast, the asymmetric initial states $\ket{01}$ and $\ket{10}$  continue to display complementary payoff distributions, preserving the symmetry under the interchange of the players and their strategies. These observations demonstrate that intermediate amplitude damping weakens, but does not completely destroy, the state-dependent quantum strategic effects, with the robustness of these effects being strongly influenced by the choice of the initial state.

\subsubsection{Maximum noise case $(p=1):$}

Under maximum amplitude damping, the quantum strategic behavior is completely suppressed and the game loses all dependence on the initial state, player strategies, and the TFIM evolution parameter. Regardless of the nature of the initial state, every strategy combination yields the identical payoff $(3,3)$ for all values of $Jt$, which is consistent with Ref.~\cite{vijayakrishnan2025influence}. Since every strategy combination leads to the same payoff, the players’ choices become irrelevant, and the quantum game collapses into a trivial classical cooperative outcome. These results demonstrate that maximum amplitude damping completely erases the quantum correlations and state-dependent strategic effects that distinguish the quantum Prisoner's Dilemma from its classical counterpart.

\section{\label{sec:level4}Dynamics of the system}

Although the payoff analysis reveals the influence of decoherence on the strategic behavior of the players, it does not directly explain the underlying evolution of quantum correlations. To gain a deeper understanding of these effects, the concurrence of the final quantum state is calculated. In particular, the focus is on the entanglement sudden birth and death under amplitude damping, and their dependence on the initial state and the TFIM evolution parameter.

\subsection{Concurrence  Measure}

To quantify the entanglement of the final two-qubit state, concurrence is employed as a widely used measure. Following Ref.~\cite{Wootters1998}, the concurrence is obtained from
\begin{equation}
    R=\rho_f(\sigma_y \otimes\sigma_y)\rho_f^{*}(\sigma_y \otimes\sigma_y)
\end{equation}
and is given by
\begin{equation}
C=\max\{0,\lambda_1-\lambda_2-\lambda_3-\lambda_4\}
\end{equation}
where $\lambda_i$ are the square roots of the eigenvalues of $R$ arranged in decreasing order.\\
The concurrence is evaluated for different initial states under the classical strategy combinations. The result reveals on the sensitivity of the TFIM-generated quantum correlations on to state preparation, and whether the observed quantum advantage and entanglement dynamics are universal or configuration-dependent. It is observed that non-zero concurrence is generated only for the asymmetric strategy pairs $(I,\sigma_x)$ and $(\sigma_x,I)$, irrespective of the choice of the initial state.
 For each of the symmetric initial states, $\ket{00}$ and $\ket{11}$, these two strategy combinations yield identical concurrence profiles. However, the concurrence evolution differs between the two initial states, indicating that the equivalence is preserved only within each initial state. For the asymmetric initial states, a complementary equivalence is observed: the concurrence obtained for $\ket{01}$ with $(I, \sigma_x)$ coincides with that of $\ket{10}$ under $(\sigma_x,I)$ , whereas $\ket{01}$ with $(\sigma_x,I)$ reproduces the concurrence profile of $\ket{10}$ with $(I, \sigma_x)$.\\
 When there is no decoherence, concurrence of the final state obtained through the modified EWL game protocol is,
 \begin{equation}
     C=|\sin(2Jt)|.
 \end{equation}
 This result represents the entanglement-generating capability of the TFIM interaction in the absence of damping. 
 \FloatBarrier

\subsection{Entanglement Dynamics of Symmetric Initial States}

\subsubsection{Initial state $\ket{00}:$}

\begin{figure}[!h]
\includegraphics[width=\columnwidth]{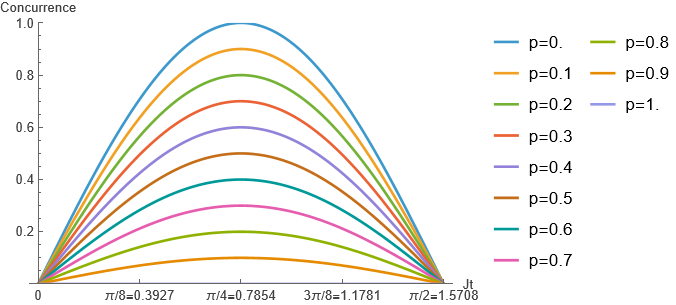}
    \caption{Concurrence as a function of the TFIM evolution parameter $Jt$ for different amplitude damping probabilities $p$, with the initial state $\ket{00}$.}
    \label{fig:00IX}
\end{figure}

The concurrence dynamics for the initial state $\ket{00}$ are presented in FIG.~\ref{fig:00IX}.
Under amplitude damping, concurrence is obtained from its corresponding final density matrix as,
\begin{equation}
C=2~max(0,|\rho_{23}|-\sqrt{\rho_{11}\rho_{44}})
   \label{eq:conc11}
\end{equation}
where $\rho_{ij}$ corresponds to the elements of the density matrix.\\
On substituting the corresponding values, Eqn.~\ref{eq:conc11} becomes,
\begin{equation}
\begin{aligned}
C &= 2\max\Bigg(0,\,
(1-p)
\left|\cos(Jt)\sin(Jt)\right|,-\sqrt{(1-p)^2\cos^2(Jt)\sin^2(Jt)}
\Bigg)\\&=(1-p)|sin(2Jt)|
\end{aligned}
\label{eq:conc000}
\end{equation}
For all degrees of noise, the concurrence increases from zero, reaches its maximum at $Jt=\pi/4$, and decreases symmetrically to zero at $Jt=\pi/2$, reflecting the entangling nature of the TFIM evolution. The maximum concurrence $(C_{\max})$ and the corresponding $Jt$ range $(\Delta Jt)$ over which it is attained, for each noise value $p$, are shown in Table~\ref{tab:cmax00}.
\begin{table}[!h]
\caption{$C_{\max}$ as a function of $p$ and their corresponding $\Delta Jt$.}
\centering
\begin{tabular}{ccc}
\hline
$p$ & $C_{\max}$ & $\Delta Jt$ \\
\hline
0.0000 & 1.0000 & 1.5707\\
0.1000 & 0.9000 & 1.5707 \\
0.2000 & 0.8000 & 1.5707\\
0.3000 & 0.7000 & 1.5707\\
0.4000 & 0.6000 & 1.5707\\
0.5000 & 0.5000 & 1.5707\\
0.6000 & 0.4000 & 1.5707\\
0.7000 & 0.3000 & 1.5707\\
0.8000 & 0.2000 & 1.5707\\
0.9000 & 0.1000 & 1.5707\\
1.0000 & 0.0000 & 1.5707\\
\hline
\end{tabular}
\label{tab:cmax00}
\end{table}
As the damping probability increases, the maximum concurrence decreases monotonically, indicating the progressive loss of entanglement. However, the concurrence remains non-zero throughout the interval $0<Jt<\pi/2$ and vanishes only at the boundaries. Thus, for the initial state $\ket{00}$, amplitude damping suppresses the entanglement continuously without exhibiting any entanglement sudden birth or death.
\FloatBarrier

\subsubsection{Initial state $\ket{11}:$}

\begin{figure}[t]
\includegraphics[width=\columnwidth]{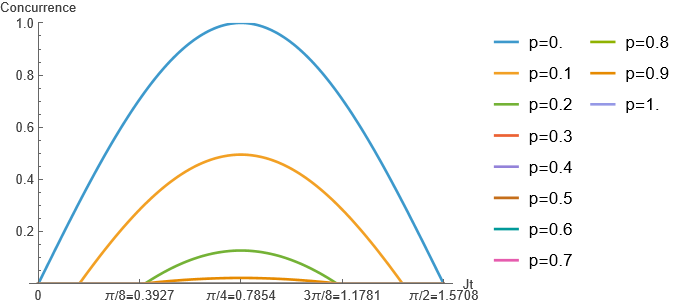}
    \caption{Concurrence as a function of the TFIM evolution parameter $Jt$ for different amplitude damping probabilities $p$, with the initial state $\ket{11}$.}
    \label{fig:11IX}
\end{figure}

FIGURE~\ref{fig:11IX} presents the concurrence for the initial state $\ket{11}$ under different amplitude-damping probabilities, with the corresponding quantitative results summarized in TABLE~\ref{tab:cmax11}. Similar to initial state $\ket{00}$, the concurrence increases from zero, reaches its maximum value at $Jt=\pi/4$, and decreases symmetrically to zero at $Jt=\pi/2$ for $p=0$. As the damping probability increases, the concurrence profile is progressively suppressed, with both the peak concurrence $C_{max}$ and the interval $\Delta Jt$ over which entanglement survives become smaller. 
\begin{table}[t]
\caption{Maximum concurrence $C_{\max}$ as a function of $p$ and their corresponding $\Delta Jt$.}
\centering
\begin{tabular}{ccc}
\hline
$p$ & $C_{\max}$ & $\Delta Jt$ \\
\hline
0.0000 & 1.0000 & 1.5707\\
0.1000 & 0.4955 & 1.2537 \\
0.2000 & 0.1281 & 0.7477\\
0.3000 & 0.0000 & 0.0000\\
0.4000 & 0.0000 & 0.0000\\
0.5000 & 0.0000 & 0.0000\\
0.6000 & 0.0000 & 0.0000\\
0.7000 & 0.0000 & 0.0000\\
0.8000 & 0.0000 & 0.0000\\
0.9000 & 0.0231 & 0.7787\\
1.0000 & 0.0000 & 0.0000\\
\hline
\end{tabular}
\label{tab:cmax11}
\end{table}
\\For damping probabilities exceeding $p\approx0.2$, the concurrence vanishes, but, interestingly, at $p\approx0.9$, a weak entanglement is observed again, indicating that a small amount of entanglement is regenerated despite the strong decoherence. For a particular damping strength, concurrence remains zero over a finite interval of $Jt$, after which it re-emerges at a specific evolution parameter $Jt_{SB}$, remains non-zero over an interval $\Delta Jt$ and subsequently becomes zero again beyond a second evolution parameter $Jt_{SD}$. Here, $Jt_{SB}$ and $Jt_{SD}$ denote the TFIM evolution parameters at which the sudden birth and sudden death of entanglement occurs, respectively. Thus, entanglement survives in the range,
\begin{equation}
    Jt_{SB}<Jt<Jt_{SD}.
\end{equation}
 Concurrence is obtained under amplitude damping using Eqn.~\ref{eq:conc11} as,
\begin{equation}
\begin{aligned}
C = 2\max\Bigg(0,\,(1-p)\left|1-2p\right|
\left|\cos(Jt)\sin(Jt)\right|-p\sqrt{(2-p)(1-p)^3(1+p(p-1))}
\Bigg)
\end{aligned}
\label{eq:conc111}
\end{equation}
Concurrence exists only when,
\begin{equation}
    |\sin(2Jt)|>S(p)
    \label{eq:ineq}
\end{equation}
where 
\begin{equation}
   S(p)= \frac{2p\sqrt{(2-p)(1-p)^3(1+p(p-1))}}{(1-p)|1-2p|};
\end{equation}
for $p \neq 1$ and $p\neq \frac{1}{2}$.\\
The inequality in Eqn.~\ref{eq:ineq} shows that the TFIM-generated entangling capability must exceed the decoherence threshold imposed by the environment.
Maximum concurrence can be obtained when $|\sin(2Jt)|=1$ which gives $Jt=\pi/4$. Thus, for all possible values of $p$, the concurrence becomes maximum at $Jt=\pi/4$. Every concurrence curve shows its maximum at $Jt=\pi/4$ as can be seen in FIG.~\ref{fig:11IX}. Since $0\le|\sin(2Jt)|\le 1$, Eqn.~\ref{eq:ineq} is satisfied only for $0\le S(p)< 1$. This is the region where entanglement can survive. The critical noise is found to be at $p=0.2451$ and $p=0.8380$ under the condition $S(p)=1$.  $S(p)<1$ is satisfied for $0\le p\le 0.2451$ and for $0.8380 < p<1$.\\
For $0\le S(p)< 1$, $sin(2Jt)=S(p)$ has two solutions which gives,
\begin{equation}
   \frac{1}{2}sin^{-1}(S(p))<Jt<\frac{1}{2}(\pi-sin^{-1}(S(p))
\end{equation}
where $ \frac{1}{2}sin^{-1}(S(p))$ is the $Jt_{SB}$ and $\frac{1}{2}(\pi-sin^{-1}(S(p))$ is the $Jt_{SD}$.
Thus, width of the Interaction-Time interval is,
\begin{equation}
    \Delta Jt=\frac{\pi}{2}-sin^{-1}(S(p))
\end{equation}
The numerical values in TABLE~\ref{tab:cmax11} can be obtained from these analytical expressions.
\FloatBarrier

\subsection{Entanglement Dynamics of Asymmetric Initial States}

All the four combinations including the  initial states $\ket{01}$ and $\ket{10}$ with the strategy pairs $(I,\sigma_x)$ and $(\sigma_x,I)$ give the same maximum concurrence, $\Delta Jt$ and same critical noise values where entanglement disappears ($p>0.4$) and reappears ($p<0.6$). These values are summarized in TABLE~\ref{tab:cmax01X}.
\begin{table}[!h]
\caption{Maximum concurrence $C_{\max}$ as a function of $p$ and their corresponding $\Delta Jt$.}
\centering
\begin{tabular}{ccc}
\hline
$p$ & $C_{\max}$ & $\Delta Jt$\\
\hline
0.0000 & 1.0000 & 1.5707\\
0.1000 & 0.6433 & 1.4248 \\
0.2000 & 0.3669 & 1.2625\\
0.3000 & 0.1630 & 1.0574\\
0.4000 & 0.0291 & 0.6885\\
0.5000 & 0.0000 & 0.0000\\
0.6000 & 0.0072 & 0.4850\\
0.7000 & 0.0400 & 0.8553\\
0.8000 & 0.0580 & 1.0523\\
0.9000 & 0.0503 & 1.2323\\
1.0000 & 0.0000 & 0.0000\\
\hline
\end{tabular}
\label{tab:cmax01X}
\end{table}
\FloatBarrier

\subsubsection{Initial state $\ket{01}$: strategy pair $(I,\sigma_x)$ (equivalent to $\ket{10}$ with $(\sigma_x,I)):$}

\begin{figure}
\includegraphics[width=\columnwidth]{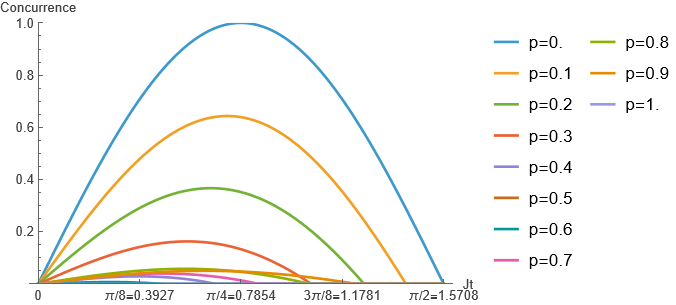}
    \caption{Concurrence as a function of the TFIM evolution parameter $Jt$ for different amplitude damping probabilities $p$, with the initial state $\ket{01}$ and strategy pair $(I,\sigma_x)$.}
    \label{fig:01IX}
\end{figure}

Initial state $\ket{01}$ with strategy pair $(I,\sigma_x)$ and initial state $\ket{10}$ with strategy pair $(\sigma_x,I)$ form the same concurrence curves as shown in FIG.~\ref{fig:01IX}.
\\Concurrence of the final state is found to be
\begin{equation}
\begin{aligned}
C= 2\max\left(0,
|\rho_{23}|-\sqrt{\rho_{11}\rho_{44}},
|\rho_{14}|-\sqrt{\rho_{22}\rho_{33}}
\right)= 2\max(0,T_1,T_2)
\end{aligned}
\label{eq:T}
\end{equation}
\vspace{-0.5em}
where
\begin{equation}
T_1= (1-p)\sin(Jt)
\Bigg[
p\cos(Jt)
\nonumber \quad
-\sqrt{(1-p)
\Big[
p^2+(1-p)
\big(
\cos^2(Jt)+p^2\sin^2(Jt)
\big)
\Big]}
\Bigg],
\end{equation}
\begin{equation}
T_2
= (1-p)\sin(Jt)
\Bigg[
-\frac{1}{2}
\Bigg(
-2(1-p)\cos(Jt)
\qquad
+\sqrt{2p}\sqrt{(2-p)
\big(
2-p+p\cos(2Jt)
\big)}
\Bigg)
\Bigg].
\end{equation}
$T_2>T_1$ for $0\le p<0.5$ and $T_2>0$ for $0\le p<0.445$ and this is the region where entanglement survives and the corresponding interaction-time interval is
\begin{align*}
0<Jt<
2\arctan(x);
\end{align*}
where 
\begin{align*}
x=\sqrt{
\frac{1-2p+3p^2-5p^3+2p^4}
{1-2p-p^2+p^3}
}
\nonumber-2\sqrt{
\frac{2p^2-7p^3+9p^4-9p^5+9p^6-5p^7+p^8}
{(1-2p-p^2+p^3)^2}
}
\end{align*}
Similarly, $T_1>T_2$ for $0.5< p<1$ and $T_1>0$ for $0.570< p<1$ and this is the region where entanglement survives and the corresponding interaction-time interval is 
\begin{align*}
0<Jt<
2\arctan(x_1);
\end{align*}
where
\begin{equation*}
x_1=\sqrt{Y-2Z}.
\end{equation*}
and 
\begin{equation*}
Y=
\frac{-1+2p+3p^2-5p^3+2p^4}
{-1+2p-p^2+p^3},~Z=
\frac{-2p^2+7p^3-5p^4-5p^5+9p^6-5p^7+p^8}
{\left(-1+2p-p^2+p^3\right)^2}
\end{equation*}
For this combination of initial states and strategic pairs, under a specific degree of noise, concurrence is non-zero as evolution happens from $Jt=0$, and hence no sudden birth occurs. As the TFIM evolution proceeds, concurrence vanishes at $Jt_{SD}$.
\FloatBarrier

\subsubsection{Initial state $\ket{01}$: strategy pair $(\sigma_x,I)$ (equivalent to $\ket{10}$ with $(I,\sigma_x)):$}

\begin{figure}[!h]
\includegraphics[width=\columnwidth]{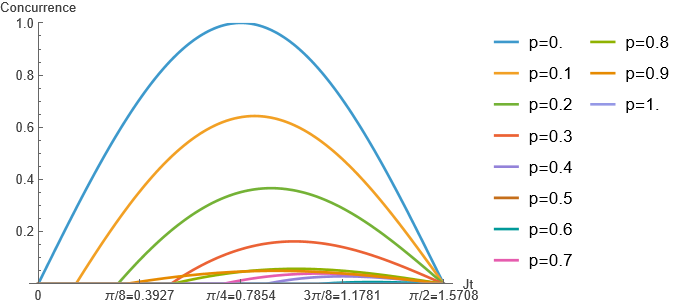}
    \caption{Concurrence as a function of the TFIM evolution parameter $Jt$ for different amplitude damping probabilities $p$, with the initial state $\ket{01}$ and strategy pair $(\sigma_x,I)$.}
    \label{fig:01IX2}
\end{figure}

Initial state $\ket{01}$ with strategy pair $(\sigma_x,I)$ and initial state $\ket{10}$ with strategy pair $(I,\sigma_x)$ form the same concurrence curves as given in FIG.~\ref{fig:01IX2}.
\\Concurrence of the final state can be calculated using \ref{eq:T},
\begin{equation}
\begin{aligned}
C = 2\max(0,T_1^{'},T_2^{'})
\end{aligned}
\end{equation}
\vspace{-0.5em}
where
\begin{equation}
T_1^{'}= -\cos(Jt)
\Bigg[
(-1+p)p\sin(Jt)
\nonumber+ \sqrt{
(-1+p)^3
\Big[
p^2\big(-1+(-1+p)\cos^2(Jt)\big)
+(-1+p)\sin^2(Jt)
\Big]
}
\Bigg],
\end{equation}
\begin{equation}
T_2^{'}= \frac{1}{2}(-1+p)\cos(Jt)
\Bigg[
\sqrt{2}\,p\sqrt{-2+p}
\sqrt{-2+p+p\cos(2Jt)}
\nonumber \quad
+2(-1+p)\sin(Jt)
\Bigg].   
\end{equation}
Again, $T_2^{'}>T_1^{'}$ for $0\le p<0.5$ and $T_2^{'}>0$ for $0\le p<0.445$ and this is the region where entanglement survives, but the corresponding interaction-time interval is
\begin{align}
2\arctan(x_0)< Jt < \frac{\pi}{2}
\label{eq:interval}
\end{align}
where $x_0$ is the positive real root of 
\begin{equation}
\begin{aligned}
(2p^2-3p^3+p^4)x_0^4+(-4+8p+2p^3-2p^4)x_0^2+(2p^2-3p^3+p^4)=0.
\label{eq:x_0}
\end{aligned}
\end{equation}
A detailed derivation of the interaction-time interval is given in Appendix~\ref{app:derivation}.\\
Similarly, $T_1^{'}>T_2^{'}$ for $0.5< p<1$ and $T_1^{'}>0$ for $0.570< p<1$ and this is the region where entanglement survives and the corresponding interaction-time interval is
\begin{align*}
2\arctan(x_1)<Jt<\pi/2.
\end{align*}
Here, under a particular damping strength, concurrence emerges at $Jt_{SB}$ and independent of all noise strengths, it becomes zero at $Jt=\pi/2$, and hence no sudden death occurs.
\FloatBarrier

\subsection{Entanglement Sudden Birth and Death}

The concurrence analysis demonstrates that the entanglement dynamics are strongly dependent on both the choice of the initial state and the classical strategy pair in an open system. For the initial state $\ket{00}$, concurrence is non-zero for $0<Jt<\pi/2$ and consequently, there is no entanglement sudden birth or sudden death. Here, the concurrence decreases monotonically with increasing damping probability for a fixed $Jt$. Thus, the TFIM evolution remains capable of producing entanglement throughout the interaction-time trajectory, corresponding to the iSWAP-family path. Thus, no finite portion of the considered iSWAP-family evolution becomes ineffective for entanglement generation for this initial state, although the degree of entanglement generated varies continuously with $Jt$ and is suppressed by amplitude damping. In contrast, traversing the same trajectory does not guarantee entanglement generation for the initial state $\ket{11}$ in the presence of amplitude damping. Only a restricted segment, $Jt_{SB}<Jt<Jt_{SD}$ produces a surviving entangled state. It displays both entanglement sudden birth and sudden death points. For the initial states $\ket{01}$  and $\ket{10}$ with strategy pairs $(I,\sigma_x)$ and $(\sigma_x,I)$ respectively, have the non-zero concurrence region $0<Jt<Jt_{SD}$ and only sudden death can be seen. Also, for the initial states $\ket{01}$  and $\ket{10}$ with strategy pairs $(\sigma_x,I)$ and $(I,\sigma_x)$ respectively, have the entangled region $Jt_{SB}<Jt<0$ and only sudden birth can be seen. Thus, for the latter three initial states, all operations in the TFIM unitaries are not capable to produce the entanglement. In these cases, as the damping probability increases, the $\Delta Jt$ over which concurrence survives becomes progressively narrower, indicating that stronger decoherence restricts the range of TFIM operators capable of sustaining quantum correlations. The concurrence value also decreases drastically as damping increases, since noise destroys the entanglement.

\subsection{Coherence}

\begin{figure}[!h]
\centering
\includegraphics[width=0.8\columnwidth]{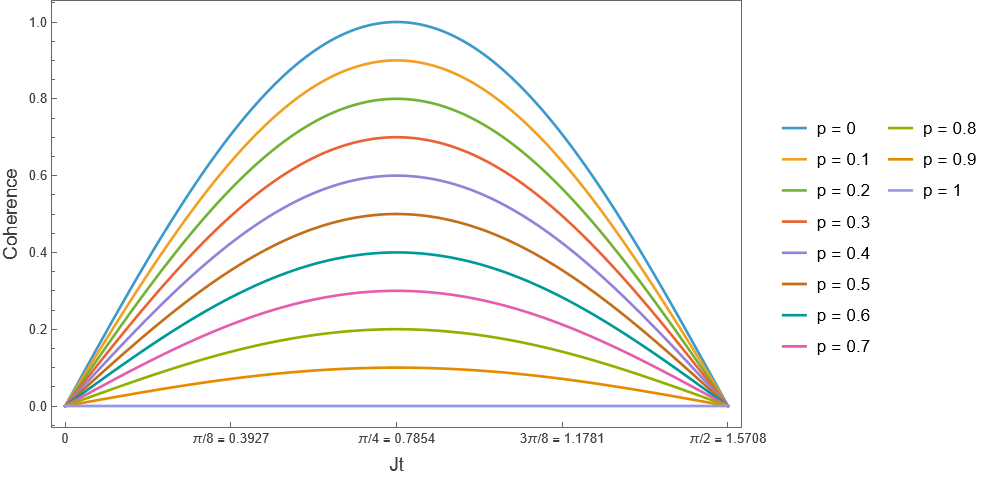}
    \caption{Coherence as a function of $Jt$ for different amplitude damping probabilities $p$, with the initial states $\ket{00}$, $\ket{01}$, $\ket{10}$}
    \label{fig:00cohIX}
\end{figure}

\begin{figure}[!h]
\centering
\includegraphics[width=0.8\columnwidth]{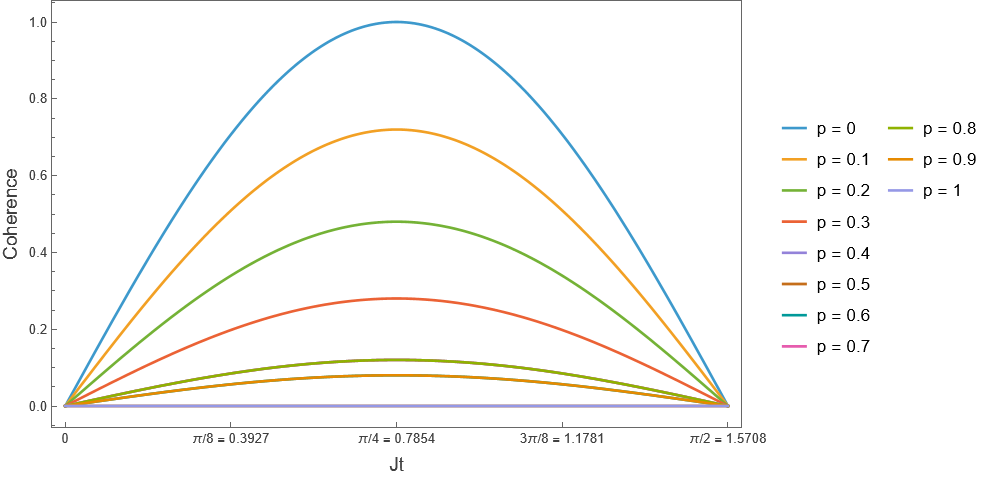}
    \caption{Coherence as a function of $Jt$ for different amplitude damping probabilities $p$, with the initial state $\ket{11}$}
    \label{fig:11cohIX}
\end{figure}

\begin{figure}[!h]
\centering
\includegraphics[width=0.6\columnwidth]{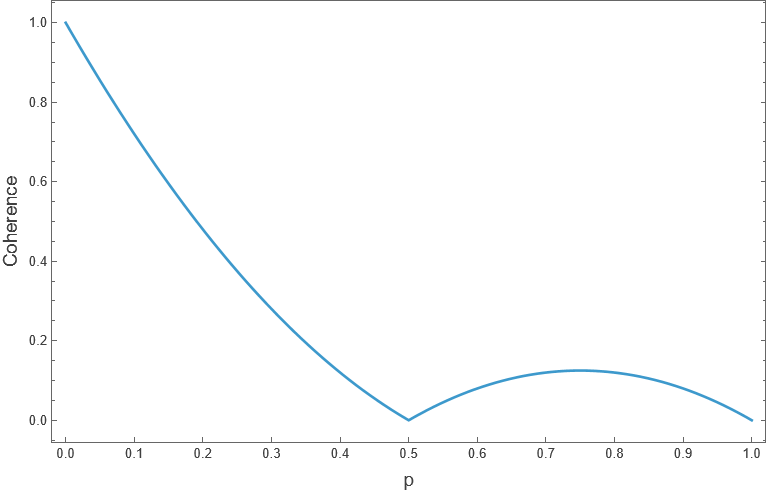}
    \caption{Coherence as a function of $Jt$ for different amplitude damping probabilities $p$, with the initial state $\ket{11}$}
    \label{fig:11cohIXp}
\end{figure}

The concurrence analysis presented above characterizes the evolution of two-qubit entanglement under amplitude damping and reveals the occurrence of entanglement sudden birth and death for specific initial states. However, entanglement and quantum coherence characterize distinct quantum resources and do not need to exhibit identical behavior under decoherence. It is entirely possible to have a system with zero entanglement that still maintains non-zero coherence~\cite{streltsov2015measuring}. On the other hand, non-zero concurrence can not exist without coherence, since entanglement is structurally built out of quantum superpositions and a state with zero coherence collapses into a purely classical mixture, immediately driving its entanglement to zero~\cite{mekala2021all}. Therefore, to obtain a more complete characterization of the effect of amplitude damping on the evolved quantum states, the quantum coherence of the final states is evaluated.\\
According to $l_1$-norm of coherence \cite{serafini2005quantifying}, quantum coherence present in the system can be quantified as,\\
\begin{equation}
    c=\sum_{i \neq j}|{\rho_f}_{ij}|
\end{equation}
Unlike concurrence, which quantifies bipartite entanglement, coherence measures the magnitude of the off-diagonal elements of the density matrix and therefore characterizes the preservation of quantum superposition in the chosen basis. Evaluating this quantity allows to independently assess how amplitude damping modifies the coherence properties of the states considered in this work and how this behavior depends on the choice of the initial state.
 \\FIGURES~\ref{fig:00cohIX} and \ref{fig:11cohIX} show that the coherence dynamics is strongly dependent on the initial state, with the states $\ket{00}$, $\ket{01}$, and $\ket{10}$ exhibiting identical coherence values, whereas $\ket{11}$ displays a distinct dependence on the damping probability.
\\Although the concurrence vanishes over certain intervals of $Jt$, nonzero coherence persists over the complete $Jt$ domain for all the initial states considered. To understand more on the dynamics, coherence is calculated for a particular evolution parameter, $Jt=\pi/4$ and is summarized in TABLE~\ref{tab:coherence}.
 \begin{table}[!h]
\caption{Coherence of the final state for $Jt=\pi/4$ as a function of the
amplitude-damping probability $p$ for different initial states.}
\centering
\begin{tabular}{ccc}
\hline
$p$ & $\ket{00}$, $\ket{01}$, $\ket{10}$ & $\ket{11}$ \\
\hline
0.0 & 1.00 & 1.00 \\
0.1 & 0.90 & 0.72 \\
0.2 & 0.80 & 0.48 \\
0.3 & 0.70 & 0.28 \\
0.4 & 0.60 & 0.12 \\
0.5 & 0.50 & 0.00 \\
0.6 & 0.40 & 0.08 \\
0.7 & 0.30 & 0.12 \\
0.8 & 0.20 & 0.12 \\
0.9 & 0.10 & 0.08 \\
1.0 & 0.00 & 0.00 \\
\hline
\end{tabular}
\label{tab:coherence}
\end{table}
\\The coherence for the initial states $\ket{00}$, $\ket{01}$, and $\ket{10}$ shows an identical monotonic decrease. But for the initial state $\ket{11}$, the coherence exhibits a non monotonic dependence on the damping probability, vanishing at $p=0.5$ and reappearing for $p>0.5$ as can be seen in FIG.~\ref{fig:11cohIXp}.
\\Thus, amplitude damping modifies the coherence generated along the TFIM evolution in a state-dependent manner, but this dependence is not directly reflected in the concurrence or its entanglement sudden birth and death structure.
\FloatBarrier

\section{\label{sec:level5}Discussion}

The present study demonstrates that the interplay between TFIM-generated quantum evolution and amplitude damping give rise to initial state and strategy-dependent entanglement dynamics in the quantum Prisoner's Dilemma.  Although the TFIM evolution follows the entanglement trajectory $(Jt,Jt,0)$ associated with the iSWAP family, entanglement is not generated uniformly across all strategy combinations. Non-zero concurrence is obtained exclusively for the asymmetric strategy pairs $(I,\sigma_x)$ and $(\sigma_x,I)$, whereas the symmetric strategy pairs remain separable throughout the evolution. This indicates that the realization of entanglement depends not only on the underlying quantum interaction but also on how the players' classical operations probe the entangling structure of the evolved operator.\\
The concurrence analysis further reveals that the initial state plays a crucial role in determining the robustness of quantum correlations against environmental decoherence. While the initial state $\ket{00}$ exhibits a gradual suppression of concurrence with increasing damping probability, it does not undergo sudden birth or death of entanglement. In contrast, the $\ket{11}$  state exhibits both sudden birth and death and the asymmetric initial states $\ket{01}$  and $\ket{10}$  exhibit either entanglement sudden birth or death depending on the strategic pair. The derived concurrence condition explains these observations by identifying the noise-dependent interaction-time interval within which entanglement can survive. As the damping probability increases, this interval becomes progressively narrower, demonstrating that stronger decoherence increasingly restricts the TFIM evolution capable of sustaining quantum correlations. In other words, the environment does not simply destroy entanglement everywhere; it restricts the $Jt$ region in which the TFIM evolution can generate and sustain it. The analytical treatment also identifies two critical damping probabilities where the entanglement disappears and re-emerges. The perfect agreement between the analytical predictions and the numerical concurrence profiles confirms the validity of the derived concurrence condition. The observation of entanglement at high damping strengths indicates that quantum correlations are not irreversibly destroyed, but can re-emerge within a limited interaction-time window due to the combined action of TFIM evolution and amplitude damping.\\
The coherence results provide a complementary perspective on the role of the TFIM evolution in the presence of amplitude damping. In contrast to concurrence, which is restricted to the interaction-time regions where bipartite entanglement is generated, the persistence of coherence over the full $Jt$ domain indicates that the TFIM evolution can maintain quantum superposition even in regimes where the final state is separable. The identical coherence behavior obtained for the $\ket{00}$, $\ket{01}$, and $\ket{10}$ initial states further indicates that, for these configurations, amplitude damping modifies the magnitude of the quantum superposition in a systematic manner that is insensitive to the particular initial state within this subset. The qualitatively different behavior of $\ket{11}$, including the temporary disappearance and subsequent recovery of coherence, shows that the effect of the channel cannot be interpreted solely as a progressive destruction of quantum resources. However, the dissipative evolution can redistribute the amplitudes and populations of the evolved state such that the off-diagonal contributions are suppressed at intermediate damping and subsequently restored over a finite range of the damping parameter. Thus, the coherence analysis demonstrates that the environmental modification of the TFIM-generated states is more nuanced than that inferred from entanglement alone. In particular, the presence of coherence outside the entangled regions shows that quantum superposition can survive even when entanglement is absent, whereas the non monotonic behavior for $\ket{11}$ highlights the sensitivity of this coherence to the combined action of the initial-state preparation and amplitude-damping channel.\\
The payoff analysis supports these entanglement and coherence results by showing that decoherence progressively suppresses the quantum advantage in the Prisoner's Dilemma. The payoff is determined from the outcome probabilities obtained from the diagonal elements of the final density matrix, whereas concurrence and coherence quantify different quantum properties of the same state, with concurrence depending on the state's entanglement structure and coherence depending on its off-diagonal elements. Under intermediate damping, the payoff retains a dependence on the initial state and the TFIM evolution, although the differences between strategic outcomes become less pronounced. Under maximum amplitude damping, all initial states and strategy combinations converge to the identical payoff $(3,3)$, rendering the players' strategic choices ineffective and reducing the game to a strategy independent outcome. This demonstrates that sufficiently strong decoherence completely masks the influence of both the TFIM-generated entanglement and the players' strategies. Also, at this limit, both concurrence and coherence vanish for all initial states and strategy pairs. Thus, complete damping simultaneously removes the quantum correlations and the strategic dependence of the game. In this work, another case where both concurrence and coherence are zero is for the initial state $\ket{11}$ with $p=0.5$. Interestingly, here also same payoff value of $(2.687,2.687)$ is obtained independent of the strategic combinations. Thus, it can be concluded that whenever this game loses its strategic dependence, its quantum correlations are also lost.\\
It is to be mentioned that a direct one-to-one correspondence between the payoff and the concurrence or coherence is not established in the present analysis. It requires a detailed investigation to establish a formal correspondence between game theoretic parameter and the physical measure. An extension of the present analysis is possible by relaxing to the condition $p_1=p_2$ and investigate the effect of asymmetric damping on the entanglement dynamics. It would also be interesting to examine other decoherence channels to determine how the characteristics of entanglement sudden birth and death depend on the nature of the system-environment interaction.\\
The overall results of this work establish that the quantum features of the TFIM-based Prisoner's Dilemma arise from the combined influence of the initial state, the classical strategy pair, and the system-environment interaction. The selective occurrence of entanglement, together with the entanglement sudden birth and death only for specific configurations, provides new insight into how decoherence modifies quantum strategic behavior. These findings will be of use in the design of robust quantum game protocols and in understanding the preservation of quantum correlations in spin-based quantum information systems.

\appendix
\section{Derivation of the interaction-time interval in Eqn.\ref{eq:interval}}
\label{app:derivation}
The $Jt$ regime for which $T_2^{'}>0$ for $0<p<0.5$ is given in Eqn.~\ref{eq:interval}. Let
\begin{equation}
A=2p^2-3p^3+p^4,
\qquad
B=-4+8p+2p^3-2p^4.
\end{equation}
Thus, Eqn.~\ref{eq:x_0} becomes,
\begin{equation}
Ax^4+Bx^2+A=0.
\end{equation}
Introducing $y=x^2$, this equation reduces to the quadratic equation
\begin{equation}
Ay^2+By+A=0.
\end{equation}
Its solutions are
\begin{equation}
y=\frac{-B\pm\sqrt{B^2-4A^2}}{2A}.
\end{equation}
For the parameter range considered here, the relevant positive real solution is
\begin{equation}
x_0=
\sqrt{\frac{-B-\sqrt{B^2-4A^2}}{2A}}.
\end{equation}
Substituting the values of $A$ and $B$ gives,
\begin{equation}
\small
x_0=
\sqrt{
\frac{
2-4p-p^3+p^4
-2\sqrt{
1-4p+4p^2-p^3+2p^4+p^5-3p^6+p^7
}
}{
p^2(p-1)(p-2)
}
}.
\end{equation}
Consequently, the lower boundary of the interaction-time interval is
\begin{equation}
Jt_{\min}=2\tan^{-1}(x_0).
\end{equation}
and the entanglement-survival interval is therefore
\begin{equation}
2\tan^{-1}(x_0)<Jt<\frac{\pi}{2}.
\end{equation}
The corresponding width of this interval can be calculated from,
\begin{equation}
\Delta Jt=\frac{\pi}{2}-2\tan^{-1}(x_0).
\end{equation}

\section*{}
\bibliographystyle{unsrt}
\bibliography{ref}

@article{nash1950equilibrium,
  title={Equilibrium points in n-person games},
  author={Nash Jr, John F},
  journal={Proceedings of the national academy of sciences},
  volume={36},
  number={1},
  pages={48--49},
  year={1950},
  publisher={national academy of sciences}
}

@book{rubinstein2007theory,
  title={Theory of Games and Economic Behavior: 60th Anniversary Commemorative Edition},
  author={Rubinstein, Ariel and Kuhn, Harold W and Morgenstern, Oskar and Von Neumann, John},
  year={2007},
  publisher={Princeton university press}
}

@article{meyer1999quantum,
  title={Quantum strategies},
  author={Meyer, David A},
  journal={Physical Review Letters},
  volume={82},
  number={5},
  pages={1052},
  year={1999},
  publisher={APS}
}

@article{eisert1999quantum,
  title={Quantum games and quantum strategies},
  author={Eisert, Jens and Wilkens, Martin and Lewenstein, Maciej},
  journal={Physical Review Letters},
  volume={83},
  number={15},
  pages={3077},
  year={1999},
  publisher={APS}
}

@article{sachdev1999quantum,
  title={Quantum phase transitions},
  author={Sachdev, Subir},
  journal={Physics world},
  volume={12},
  number={4},
  pages={33--38},
  year={1999}
}

@article{vijayakrishnan2019role,
  title={Role of two-qubit entangling operators in the modified Eisert--Wilkens--Lewenstein approach of quantization},
  author={Vijayakrishnan, V and Balakrishnan, S},
  journal={Quantum Information Processing},
  volume={18},
  number={4},
  pages={112},
  year={2019},
  publisher={Springer}
}

@article{flitney2005quantum,
  title={Quantum games with decoherence},
  author={Flitney, Adrian P and Abbott, Derek},
  journal={Journal of Physics A: Mathematical and General},
  volume={38},
  number={2},
  pages={449--459},
  year={2005}
}

@article{huang2016quantum,
  title={Quantum games under decoherence},
  author={Huang, Zhiming and Qiu, Daowen},
  journal={International Journal of Theoretical Physics},
  volume={55},
  number={2},
  pages={965--992},
  year={2016},
  publisher={Springer}
}

@article{chen2003quantum,
  title={Quantum prisoner dilemma under decoherence},
  author={Chen, LK and Ang, Huiling and Kiang, D and Kwek, LC and Lo, CF},
  journal={Physics Letters A},
  volume={316},
  number={5},
  pages={317--323},
  year={2003},
  publisher={Elsevier}
}

@article{Wootters1998,
  author  = {Wootters, William K.},
  title   = {Entanglement of Formation of an Arbitrary State of Two Qubits},
  journal = {Physical Review Letters},
  volume  = {80},
  number  = {10},
  pages   = {2245--2248},
  year    = {1998},
  doi     = {10.1103/PhysRevLett.80.2245}
}

@article{flitney2002introduction,
  title={An introduction to quantum game theory},
  author={Flitney, Adrian P and Abbott, Derek},
  journal={Fluctuation and Noise Letters},
  volume={2},
  number={04},
  pages={R175--R187},
  year={2002},
  publisher={World Scientific}
}

@article{marinatto2000quantum,
  title={A quantum approach to static games of complete information},
  author={Marinatto, Luca and Weber, Tullio},
  journal={Physics Letters A},
  volume={272},
  number={5-6},
  pages={291--303},
  year={2000},
  publisher={Elsevier}
}

@article{sarkar2019quantum,
  title={Quantum {Nash} equilibrium in the thermodynamic limit},
  author={Sarkar, Shubhayan and Benjamin, Colin},
  journal={Quantum Information Processing},
  volume={18},
  number={4},
  pages={122},
  year={2019},
  publisher={Springer}
}

@article{adami2018thermodynamics,
  title={Thermodynamics of evolutionary games},
  author={Adami, Christoph and Hintze, Arend},
  journal={Physical Review E},
  volume={97},
  number={6},
  pages={062136},
  year={2018},
  publisher={APS}
}

@article{benjamin2020emergence,
  title={The emergence of cooperation in the thermodynamic limit},
  author={Benjamin, Colin and Sarkar, Shubhayan},
  journal={Chaos, Solitons \& Fractals},
  volume={135},
  pages={109762},
  year={2020},
  publisher={Elsevier}
}

@article{galam2010ising,
  title={Ising model versus normal form game},
  author={Galam, Serge and Walliser, Bernard},
  journal={Physica A: Statistical Mechanics and its Applications},
  volume={389},
  number={3},
  pages={481--489},
  year={2010},
  publisher={Elsevier}
}

@article{fan2005optimal,
  title={Optimal two-qubit quantum circuits using exchange interactions},
  author={Fan, Heng and Roychowdhury, Vwani and Szkopek, Thomas},
  journal={Physical Review A—Atomic, Molecular, and Optical Physics},
  volume={72},
  number={5},
  pages={052323},
  year={2005},
  publisher={APS}
}

@book{nielsen2000quantum,
  title={Quantum computation and quantum information},
  author={Nielsen, Michael A and Chuang, Isaac L and others},
  volume={1},
  year={2000},
  publisher={Cambridge university press Cambridge}
}

@article{piotrowski2003invitation,
  title={An invitation to quantum game theory},
  author={Piotrowski, Edward W and S{\l}adkowski, Jan},
  journal={International Journal of Theoretical Physics},
  volume={42},
  number={5},
  pages={1089--1099},
  year={2003},
  publisher={Springer}
}

@article{vijayakrishnan2025influence,
  title={Influence of noise on a quantum game in the light of modified EWL scheme},
  author={Vijayakrishnan, V and Balakrishnan, S},
  journal={Quantum Information Processing},
  volume={24},
  number={7},
  pages={203},
  year={2025},
  publisher={Springer}
}

@article{makhlin2002nonlocal,
  title={Nonlocal properties of two-qubit gates and mixed states, and the optimization of quantum computations},
  author={Makhlin, Yuriy},
  journal={Quantum Information Processing},
  volume={1},
  number={4},
  pages={243--252},
  year={2002},
  publisher={Springer}
}

@article{rezakhani2004characterization,
  title={Characterization of two-qubit perfect entanglers},
  author={Rezakhani, AT},
  journal={Physical Review A},
  volume={70},
  number={5},
  pages={052313},
  year={2004},
  publisher={APS}
}

@article{zhang2003geometric,
  title={Geometric theory of nonlocal two-qubit operations},
  author={Zhang, Jun and Vala, Jiri and Sastry, Shankar and Whaley, K Birgitta},
  journal={Physical Review A},
  volume={67},
  number={4},
  pages={042313},
  year={2003},
  publisher={APS}
}

@article{yu2005evolution,
  title={Evolution from entanglement to decoherence of bipartite mixed" X" states},
  author={Yu, Ting and Eberly, Joseph H},
  journal={arXiv preprint quant-ph/0503089},
  year={2005}
}

@article{yu2009sudden,
  title={Sudden death of entanglement},
  author={Yu, Ting and Eberly, Joseph H},
  journal={Science},
  volume={323},
  number={5914},
  pages={598--601},
  year={2009},
  publisher={American Association for the Advancement of Science}
}

@article{yang2013relation,
  title={Relation between initial conditions and entanglement sudden death for two-qubit extended Werner-like states},
  author={Yang, Bai-Yuan and Fang, Mao-Fa and Huang, Jiang},
  journal={Chinese Physics B},
  volume={22},
  number={8},
  pages={080303},
  year={2013}
}

@article{yu2006sudden,
  title={Sudden death of entanglement: classical noise effects},
  author={Yu, Ting and Eberly, JH},
  journal={Optics Communications},
  volume={264},
  number={2},
  pages={393--397},
  year={2006},
  publisher={Elsevier}
}

@article{lopez2008sudden,
  title={Sudden birth versus sudden death of entanglement in multipartite systems},
  author={L{\'o}pez, CE and Romero, G and Lastra, F and Solano, E and Retamal, JC},
  journal={Physical Review Letters},
  volume={101},
  number={8},
  pages={080503},
  year={2008},
  publisher={APS}
}

@article{xu2009sudden,
  title={Sudden death and birth of entanglement beyond the Markovian approximation},
  author={Xu, ZY and Feng, M},
  journal={Physics Letters A},
  volume={373},
  number={22},
  pages={1906--1910},
  year={2009},
  publisher={Elsevier}
}

@article{serafini2005quantifying,
  title={Quantifying decoherence in continuous variable systems},
  author={Serafini, Alessio and Paris, Matteo GA and Illuminati, Fabrizio and De Siena, Silvio},
  journal={Journal of Optics B: Quantum and Semiclassical Optics},
  volume={7},
  number={4},
  pages={R19--R36},
  year={2005}
}

@article{streltsov2015measuring,
  title={Measuring quantum coherence with entanglement},
  author={Streltsov, Alexander and Singh, Uttam and Dhar, Himadri Shekhar and Bera, Manabendra Nath and Adesso, Gerardo},
  journal={Physical review letters},
  volume={115},
  number={2},
  pages={020403},
  year={2015},
  publisher={APS}
}

@article{mekala2021all,
  title={All entangled states are quantum coherent with locally distinguishable bases},
  author={Mekala, Asmitha and Sen, Ujjwal},
  journal={Physical Review A},
  volume={104},
  number={5},
  pages={L050402},
  year={2021},
  publisher={APS}
}

@article{Thomas2026StrategicEquilibria,
author  = {Thomas, Teena and Balakrishnan, S.},
title   = {Emergence of Strategic Equilibria from Transverse Field {Ising} Hamiltonian Dynamics},
journal = {arXiv preprint arXiv:2608.29926},
year    = {2026},
eprint  = {2608.29926},
archivePrefix = {arXiv},
primaryClass = {quant-ph}
}

\end{document}